\documentstyle[11pt,newpasp,twoside,epsfig]{article}
\def\edcomment#1{\iffalse\marginpar{\raggedright\sl#1\/}\else\relax\fi}
\reversemarginpar

\begin{document}
\title{Chemical Evolution of Elliptical Galaxies as a Constraint to Galaxy 
       Formation Scenarios}
\author{Donatella Romano}
\affil{International School for Advanced Studies, SISSA/ISAS, 
       Via Beirut 2-4, I-34014 Trieste, Italy}
\author{Francesca Matteucci}
\affil{Dipartimento di Astronomia, Universit\`a di Trieste,
       Via G.B. Tiepolo 11, I-34131 Trieste, Italy}
\author{Luigi Danese}
\affil{International School for Advanced Studies, SISSA/ISAS, 
       Via Beirut 2-4, I-34014 Trieste, Italy}

\begin{abstract}
Elliptical galaxies are the main contributors to the chemical enrichment of 
the intracluster and intergalactic medium; understanding how they form and 
evolve enables us to get important hints on the amounts of energy and 
processed matter that they eject into the ICM/IGM. Recent pieces of 
observational evidence point to a strong connection between high redshift 
quasars and their host galaxies. The aim of this paper is to prove that the 
main aspects of the chemical evolution of the spheroids can be reproduced in 
the framework of a model where the shining of the quasar is intimately related 
to the formation of the galactic nucleus. A key assumption is that the quasars 
shone in an inverted order with respect to the hierarchical one (i.e., stars 
and black holes in bigger dark halos formed before those in smaller ones) 
during an early episode of vigorous star formation. This scenario closely 
resembles the so-called `inverse wind' model invoked to explain the observed 
increase of the [Mg/Fe] ratio in the nuclei of ellipticals with increasing the 
galactic mass, the only difference being that now the time for the occurrence 
of a galactic wind is not determined by the energy input from supernovae, but 
is indeed the energy injected by the quasar which regulates the onset of the 
wind phase.
\end{abstract}

\section{Introduction}
Most of the information on elliptical galaxies comes from their integrated 
properties: by analyzing colors and spectra by means of population synthesis 
techniques (e.g., Buzzoni et al. 1992; Bruzual \& Charlot 1993; Bressan et al. 
1994, 1996; Tantalo et al. 1998) one can infer the real abundances of the 
dominant stellar populations. A value of the elemental ratio [Mg/Fe] $>$ 0 in 
the nuclei of ellipticals has been derived; moreover, [Mg/Fe] has been found 
to increase with increasing the galactic mass (Faber et al. 1992; Worthey et 
al. 1992; Weiss et al. 1995). The overabundance of [Mg/Fe] with respect to 
solar is generally interpreted as due to a short and intense star formation, 
perhaps coupled to an IMF slightly biased towards massive stars. Other 
observational hints favoring a scenario in which the bulk of the stellar 
population is build up on a short timescale are the existence of a tight 
fundamental plane (FP) in the 3-space of the basic global parameters central 
velocity dispersion $\sigma$, effective radius $r_e$, and mean effective 
surface brightness $I_e$ (Bender et al. 1992; Renzini \& Ciotti 1993) and the 
tight color--$\sigma$ and color--magnitude relations (Bower et al. 1992) for 
ellipticals in local clusters, and the modest shift with redshift in the 
zero-point of the FP, Mg$_2$--$\sigma$, and color--magnitude relations of 
cluster ellipticals at intermediate $z$ (Dickinson 1995; Ellis et al. 1997; 
Bender et al. 1998; van Dokkum et al. 1998; Kodama et al. 1998; Stanford et 
al. 1998). A maximum age difference of $\sim$ 1 Gyr between the bulk of the 
stellar population in field and cluster ellipticals at given mass has been 
inferred (Bernardi et al. 1998; see also Concannon et al. 2000 and Maraston \& 
Thomas 2000), thus suggesting that most stars in ellipticals formed at $z$ $>$ 
3, independently on the environment. The present day Type Ia and II SN rates 
in early-type galaxies ($Rate_{SNIa}$ = 0.18 $\pm$ 0.06 SNu, $Rate_{SNII}$ $<$ 
0.02 SNu for $H_0$ = 75 km s$^{-1}$ Mpc$^{-1}$; Cappellaro et al. 1999) are 
another argument in favor of a scenario where a long-lasting phase of passive 
evolution follows an early episode of vigorous star formation. In fact, since 
Type II SNe come from short-living progenitors whereas Type Ia SNe come from 
long-living ones, the observed Type Ia and II SN rates imply that star 
formation in early-type galaxies must be nearly inactive at the present time. 

In recent years, a strong connection between quasi-stellar objects (QSOs) 
observed at high redshifts and their host proto-galaxies has become apparent. 
In particular, the correlation between the central massive dark object and the 
hot stellar component of nearby galaxies (e.g., Kormendy \& Richstone 1995; 
Magorrian et al. 1998; van der Marel 1999; Ferrarese \& Merritt 2000; Gebhardt 
et al. 2000) suggests a direct connection between QSO activity and galaxy 
formation. Several models taking into account jointly the cosmological 
formation of QSOs and spheroids have appeared in the literature (Silk \& Rees 
1998; Fria\c ca \& Terlevich 1998; Monaco et al. 2000; Granato et al. 2001, 
among others). Following Monaco et al. (2000) and Granato et al. (2001), the 
QSOs shine at the centers of the spheroids while a vigorous star formation is 
building up the bulk of the stellar population. As a consequence of this, the 
surrounding medium is ionized, the star formation stops and a galactic wind 
develops. The QSOs shine in an {\it inverted hierarchical order} (Monaco et 
al. 2000), i.e., more massive spheroids experience the shining of the QSO 
before less massive ones. This scenario closely resembles the `inverse wind' 
scenario invoked to explain the increase of the [Mg/Fe] ratio of the stellar 
populations in the nuclei of ellipticals with increasing total galactic 
luminosity (Matteucci 1994). We will show that by adopting the QSO shining 
times given by Monaco et al. (2000) and Granato et al. (2001) we can account 
for the main chemical properties of the stellar populations in early-type 
galaxies, thus giving further support to their model, which already 
reproduces: i) the evolution of the quasar luminosity function, ii) the 850 
and 450 $\mu$m source counts together with their related statistics, iii) the 
mass function of dormant black holes in nearby galaxies and iv) the 
correlation of the black hole mass with the mass of the host galaxy spheroid.

\section{The Model}

Elliptical galaxies are considered initially as spheres of gas with luminous 
mass in the range $M_{gas}$ $\sim$ 1.0 $\times$ 10$^{10}$--2.5 $\times$ 10$^
{12}$ $M_\odot$ embedded in a massive dark halo of mass $M$ $\sim$ 
7\,$M_{gas}$. At the beginning, both baryonic and dark matter follow a 
Navarro, Frenk, \& White (1997) density profile. Then, baryonic matter cools, 
collapses and starts forming stars. We assume a single zone interstellar 
medium (ISM) with instantaneous mixing of gas throughout. The star formation 
builds up the bulk of the stellar population with high efficiency in a short 
timescale. Then, the QSO shines at the centre, ionizes the ISM and stops the 
star formation. A galactic wind eventually develops at this stage. The shining 
times, $t_{QSO}$, are listed in Table 1 as a function of the initial gaseous 
mass. We compute the cooling time and the dynamical time at each radius and 
define $r_{cool}$, the maximum radius at which both the cooling time and the 
dynamical time are still lower than $t_{QSO}$. Obviously, only matter inside 
$r_{cool}$ (a fraction $\alpha$ of the initial gaseous mass) will enter the 
process of star formation. The quantity $\alpha$ is listed in Table 1.

The star formation rate (SFR) is given by:
\begin{equation}
\psi(t) = \nu M_{cold}(t),
\end{equation}
where $M_{cold}(t)$ is the gas mass which has cooled by the time $t$. The 
quantity $\nu$, expressed in units of Gyr$^{-1}$, represents the efficiency of 
star formation, namely the inverse of the timescale of star formation. The 
timescale of star formation is the maximum between the mean values of the 
cooling time and the dynamical time inside $r_{cool}$.

The fundamental equations of chemical evolution, which allow us to follow the 
temporal evolution of the abundances of several elemental species in the gas 
are:
\begin{equation}
\frac{{\mathrm d}G_i(t)}{{\mathrm d}t} = - X_i(t) \psi(t) + R_i(t) + 
\Big( \frac{{\mathrm d}G_i}{{\mathrm d}t} \Big)_{infall} - 
\Big( \frac{{\mathrm d}G_i}{{\mathrm d}t} \Big)_{reheat}.
\end{equation}
$G_i(t)$ = $X_i(t) M_{cold}(t)$ is the cold gas mass in form of the element 
$i$; $X_i(t)$ is the abundance by mass of the element $i$; $R_i(t)$ is the 
rate at which dying stars eject both processed and unprocessed matter (see 
Matteucci \& Greggio 1986). The last two terms on the right account for the 
accretion of cold gas by infall and for the reheating due to supernova 
explosions, namely the rate at which the gas is heated by SNe and therefore 
subtracted to star formation.

As far as the initial mass function (IMF) is concerned, we choose either a 
Salpeter IMF, $\phi(M)$ $\propto$ $M^{-1.35}$, or $\phi(M)$ $\propto$ 
$M^{-1.15}$. The normalization is performed in the mass range 0.1--100 
$M_\odot$.

The average abundances of the composite stellar populations we use are the 
mass-averaged ones, namely:
\begin{equation}
\langle X_i \rangle = \frac{1}{S_{tot}} \int_0^{S_{tot}} X_i(S) {\mathrm d}S,
\end{equation}
where $S_{tot}$ is the total mass of stars ever born.

\begin{table}
\caption{Model parameters [columns from (2) to (5)] together with some model 
results [columns from (6) to (9)]. See text for details.}
\begin{tabular}{@{} c c c c c c c c c @{}}
\multicolumn{9}{c}{}\\
\tableline
Model & $M_{gas}$ & $t_{QSO}$ & $\alpha$ & $\nu$ & $M_{stars}$ & 
$\langle$[Mg/Fe]$\rangle$ & Mg$_2$ & $\langle$Fe$\rangle$ \\
(1) & (2) & (3) & (4) & (5) & (6) & (7) & (8) & (9) \\
\tableline
1 & 2.5 $\times$ 10$^{12}$ & 0.6 & 0.15 & 4.8 & 2.7 $\times$ 10$^{11}$ & 
0.443 & 0.246 & 2.614$^{\rm a}$ \\
\vspace{0.1cm}
{} & {} & {} & {} & {} & 2.9 $\times$ 10$^{11}$ & 0.550 & 0.310 & 
2.982$^{\rm b}$ \\
2 & 9.5 $\times$ 10$^{11}$ & 0.7 & 0.20 & 4.3 & 1.6 $\times$ 10$^{11}$ & 
0.424 & 0.249 & 2.658$^{\rm a}$ \\
\vspace{0.1cm}
{} & {} & {} & {} & {} & 1.7 $\times$ 10$^{11}$ & 0.537 & 0.314 & 
3.016$^{\rm b}$ \\
3 & 5.5 $\times$ 10$^{10}$ & 1.6 & 0.70 & 1.5 & 2.4 $\times$ 10$^{10}$ & 
0.333 & 0.227 & 2.561$^{\rm a}$ \\
\vspace{0.1cm}
{} & {} & {} & {} & {} & 2.3 $\times$ 10$^{10}$ & 0.471 & 0.280 & 
2.838$^{\rm b}$ \\
4 & 2.0 $\times$ 10$^{10}$ & 2.1 & 0.90 & 1.1 & 8.0 $\times$ 10$^{9}$ & 0.288 
& 0.209 & 2.456$^{\rm a}$ \\
\vspace{0.1cm}
{} & {} & {} & {} & {} & 6.9 $\times$ 10$^{9}$ & 0.438 & 0.253 & 
2.670$^{\rm b}$ \\
5 & 1.0 $\times$ 10$^{10}$ & 2.5 & 1.00 & 1.0 & 3.5 $\times$ 10$^{9}$ & 0.249 
& 0.193 & 2.348$^{\rm a}$ \\
{} & {} & {} & {} & {} & 2.8 $\times$ 10$^{9}$ & 0.410 & 0.228 & 
2.507$^{\rm b}$ \\
\tableline
\tableline
\end{tabular}
\begin{list}{}{}
\item[$^{\rm a}$] $\phi(M)$ $\propto$ $M^{-1.35}$; $^{\rm b}$ $\phi(M)$ 
$\propto$ $M^{-1.15}$
\end{list}
\end{table}

\section{Model results and discussion}

In Table 1 we report model parameters [column (2): initial gaseous mass, in 
units of $M_\odot$; column (3): QSO shining time, in units of Gyr; column (4): 
fraction of $M_{gas}$ which cools and collapses before $t_{QSO}$; column (5): 
star formation efficiency, in units of Gyr$^{-1}$] and results [columns (6): 
mass converted into stars from $t$ = 0 to $t$ = $t_{QSO}$, in units of 
$M_\odot$; column (7): average [Mg/Fe] of the composite stellar population at 
$t$ = $t_{QSO}$; columns (8) and (9): metallicity indices Mg$_2$ and 
$\langle$Fe$\rangle$ of the composite stellar population at $t$ = $t_{QSO}$]. 
Results for both $\phi(M)$ $\propto$ $M^{-1.35}$ and $\phi(M)$ $\propto$ 
$M^{-1.15}$ are shown for each model galaxy.

Owing to the fact that the QSOs in more massive host proto-galaxies shine 
before those in less massive ones, larger $\langle$[Mg/Fe]$\rangle$ ratios are 
displayed by the stellar populations of more massive spheroids, in agreement 
with observations. In fact, a prolongued star formation period results in 
adding younger stellar components formed out of iron-enriched gas, due to the 
temporal delay in type Ia SN explosions which restore the bulk of iron (e.g., 
Matteucci 1994). Moreover, at the time of the shining of the QSO, all our 
massive galaxies have reached a roughly solar or even super-solar metallicity 
at the center, in agreement with observational hints on the metal content in 
QSO environments (Hamann \& Ferland 1999). The theoretical 
$\langle$[Mg/Fe]$\rangle$, $\langle$[Fe/H]$\rangle$ ratios can be converted 
into metallicity indices following the prescriptions by Matteucci et al. 
(1998) to be compared to observations in the Mg$_2$--$\langle$Fe$\rangle$ 
diagram. We want to stress that introducing a reheating term in Eq.(2) helps 
us in reproducing the observed range of definition of the Mg$_2$ index (see Fig.1). 
However, at present we used a very simplified treatment for the SN feedback, 
taking into account only type II SNe.

\begin{figure}
\plotone{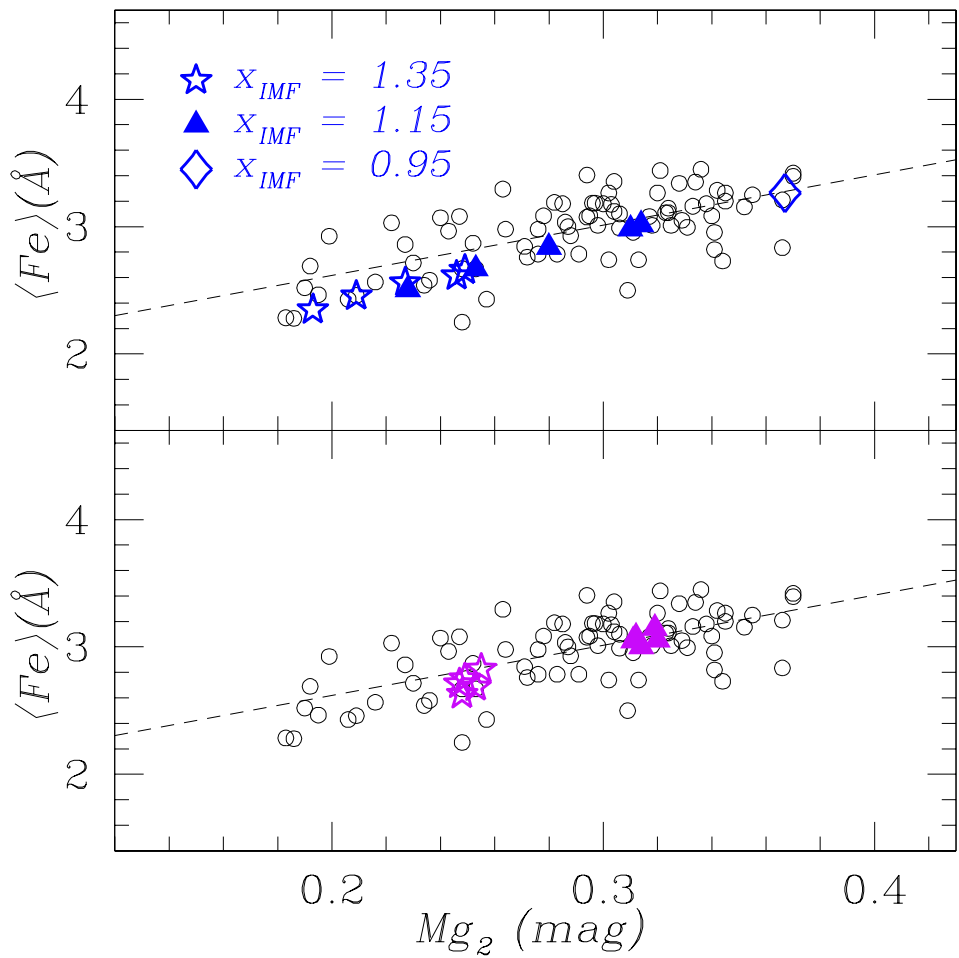}
\caption{Mg$_2$ vs. $\langle$Fe$\rangle$ theoretical relations compared to 
the available data ({\it open circles}; Worthey et al. 1992; Gonz\' alez 1993; 
Carollo \& Danziger 1994a, b). Results relevant to $\phi(M)$ $\propto$ 
$M^{-1.35}$ ({\it stars}) and to $\phi(M)$ $\propto$ $M^{-1.15}$ ({\it 
triangles}) are shown for Models from 1 to 5. The same results are listed in 
Table 1. The big diamond refers to Model 1 computed with an even steeper IMF 
($\phi(M)$ $\propto$ $M^{-0.95}$). Results in the top panel refer to models in 
which the reheating term [see Eq.(2)] is taken into account, whereas results 
in the bottom panel refer to models in which the reheating term is set to 
zero. We conclude that accounting for the reheating due to SN explosions helps 
us in reproducing the observed range of definition of the Mg$_2$ index.}
\end{figure}
\begin{figure}
\plotfiddle{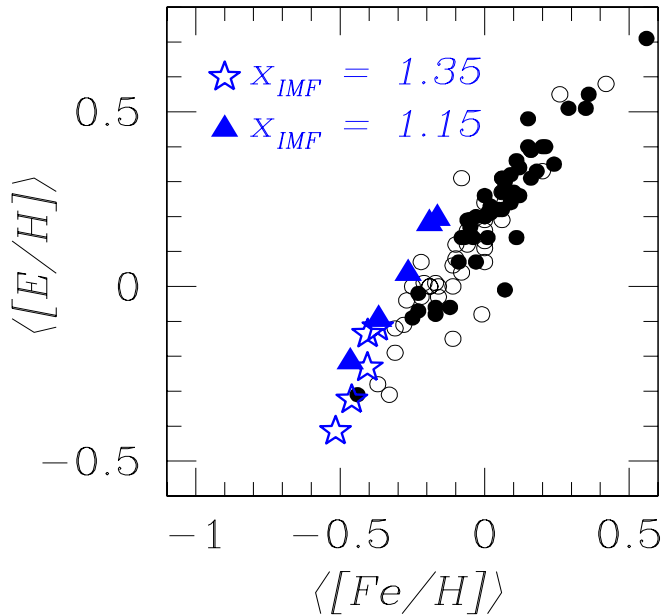}{7.5cm}{0}{45}{43}{-150}{-72}
\caption{Theoretical $\langle$[E/H]$\rangle$ vs. $\langle$[Fe/H]$\rangle$ 
compared to the data from Trager et al. 2000 (see text for details).}
\end{figure}

Recently, Trager et al. (2000) derived values of the $\langle$[E/Fe]$\rangle$ 
and $\langle$[Fe/H]$\rangle$ ratios for a sample of ellipticals from a set of 
observed indices (E refers to all `enhanced' species, see Trager et al.). In 
Fig.2 we compare our theoretical predictions to their data, relevant to both 
$r_e$/8 ({\it filled circles}) and $r_e$/2 ({\it open circles}). Since we 
adopt a one-zone model, our results should be considered as relevant to an 
aperture $r_e$. The agreement with the data is quite good. The interesting 
point is that we do not find the need for a significant young stellar 
component to be added to the older one in order to reproduce the data.

We have tested the coupled QSO-host galaxy formation scenario against the 
main chemical properties of the composite stellar populations of early-type 
galaxies and shown that they can indeed be reproduced, at least in the 
framework of the simple one-zone chemical evolution model described here. 
It has been recognised that SNe alone are unlikely to be the source of the 
overall IGM heating and need to be supplemented or even substituted by some 
other heating processes (Kravtsov \& Yepes 2000 and refs. therein). Radiation 
from QSO population could provide the required heating (Valageas \& Silk 
1999). Therefore, further study should be deserved to the topic of the 
QSO-host galaxy connection.

\end{document}